%% file: template.tex
\documentclass[preprints,article,accept,moreauthors,pdftex,a4paper]{Definitions/mdpi}
\pdfoutput=1

\firstpage{1} 
\pubvolume{xx}
\issuenum{1}
\articlenumber{5}
\pubyear{2021}
\copyrightyear{2021}
\history{Submit preprint: 01.02.2021}

\usepackage{pgfplots, siunitx}
\usepackage[ruled,vlined]{algorithm2e}
\pgfplotsset{compat=1.16}

\Title{A Novel Approach for Classification and Forecasting of Time Series in Particle Accelerators}

\Author{Sichen Li$^{1,\dagger}$\orcidA{}, M\'elissa Zacharias$^{1,\dagger}$, Jochem Snuverink$^{1}$, Jaime Coello de Portugal$^{1}$, Fernando Perez-Cruz$^{2}$, Davide Reggiani$^{1}$ and Andreas Adelmann$^{1}$*\orcidB{}}

\AuthorNames{Sichen Li, M\'elissa Zacharias, Jochem Snuverink, Jaime Coello de Portugal, Fernando Perez-Cruz, Davide Reggiani and Andreas Adelmann}

\address{Paul Scherrer Institut, 5232 Villigen PSI, Switzerland}

\address{%
$^{1}$ \quad Paul Scherrer Institut, 5232 Villigen PSI, Switzerland\\
$^{2}$ \quad Swiss Data Science Center, ETH Zürich and EPFL, Universitätstrasse 25 8092,
Zürich, Switzerland}

\corres{Correspondence: andreas.adelmann@psi.ch}
\firstnote{These authors contributed equally to this work.}

\abstract{
The beam interruptions (interlocks) of particle accelerators, despite being necessary safety measures, lead to abrupt operational changes and a substantial loss of beam time. 
A novel time series classification approach is applied to decrease beam time loss in the High Intensity Proton Accelerator complex by forecasting interlock events.
The forecasting is performed through binary classification of windows of multivariate time series. The time series are transformed into Recurrence Plots which are then classified by a Convolutional Neural Network, which not only captures the inner structure of the time series but also utilizes the advances of image classification techniques. 
Our best performing interlock-to-stable classifier reaches an Area under the ROC Curve value of \texorpdfstring{$0.71 \pm 0.01$}{0.71±0.01} compared to \texorpdfstring{$0.65 \pm 0.01$}{0.65±0.01} of a Random Forest model, and it can potentially reduce the beam time loss by \texorpdfstring{$0.5 \pm 0.2$}{0.5±0.2} seconds per interlock. 
}


\begin{document}

\section{Introduction}
\unskip
\input{1_Introduction}

\section{Materials and Methods}
\unskip
\subsection{Dataset and preprocessing}
\input{2.1_Dataset}

\subsection{Problem Formulation}\label{sec:problem}
\input{2.2_Problem}

\subsection{Recurrence Plot (RP)}
The HIPA channels present behaviors on various time scales, such as slow signal drifts or sudden jumps, which are very hard to clean up in the manner necessary for usual time series prediction methods. Recurrence plots are intrinsically suited to deal with these issues, as they are capable of extracting time dependent correlations on different time scales and complexity via tunable parameters.
    \subsubsection{Theory}
    \input{2.3.1_Recurrence_theory}

    \subsubsection{RPCNN Model}
    \input{2.3.2_Recurrence_methods}
	
\subsection{Random Forest}
\input{2.4_RF}

\subsection{Evaluation metric}
\input{2.5_Metric}

\section{Results}
\unskip
\subsection{Model performance in terms of evaluation metrics}
\input{3.1_Results_evaluation}
\subsection{Model performance in a simulated live setting}\label{sec:live}
\input{3.2_Results_live}
\section{Discussion}
\input{4_Discussion}

\vspace{6pt} 


\authorcontributions{conceptualization, S.L., F.P. and A.A.; methodology, M.Z., F.P. and A.A.; software, M.Z. and S.L.; validation, J.S., J.C. and A.A.; formal analysis, M.Z. and S.L.; investigation, J.S., J.C. and D.R.; resources, J.S. and D.R.; data curation, J.S., J.C. and D.R.; writing--original draft preparation, S.L. and M.Z.; writing--review and editing, J.S., J.C., F.P. and A.A.; visualization, S.L. and M.Z.; funding acquisition, J.S. and A.A.}

\funding{
This research was partly funded by the Swiss Data Science Center.}

\acknowledgments{We would like to place our special thanks to Dr.~Anastasia Pentina and other colleagues from the Swiss Data Science Center for their insightful collaboration and generous support throughout the research. We also thank Hubert Lutz and Simon Gregor Ebner from PSI for their expert knowledge on the HIPA Archiver and help in data collection. We acknowledge the assistance of Dr.\ Derek Feichtinger and Marc Chaubet for their help with the Merlin cluster which enables the computational work of the research. }

\conflictsofinterest{The authors declare no conflict of interest. The funders had no role in the design of the study; in the collection, analyses, or interpretation of data; in the writing of the manuscript, or in the decision to publish the results.}




\appendixtitles{no} 
\appendix
\input{Appendix}


\externalbibliography{yes}
\bibliography{template.bib}



\end{document}

%% file: 1_Introduction.tex
Recent years have seen a boost in the development of machine learning (ML) algorithms and their applications \cite{edelen2018opportunities}. With the rapid growth of data volume and processing capabilities, the value of data has been increasingly recognized both academically and in social life, and the prospect of ML has been pushed to an unprecedented level.

Particle accelerators, as large facilities operating with a steady stream of structured data, are naturally suitable for ML applications.
The highly complicated operation conditions and precise control objectives fall perfectly inside ML's scope~\cite{edelen2016neural}.
Over the past few years, the interest and engagement of ML applications in the field of particle accelerators has started to grow~\cite{arpaia2020machine, edelen2018opportunities} along with the trend of data-driven approaches in other disciplines. Fast and accurate ML-based beam dynamics modelling can serve as a guide for future accelerator design and commissioning~\cite{zhao2020beam}. An ML surrogate model could achieve the same precision by utilizing only a few runs of the original high fidelity simulation~\cite{adelmann2019nonintrusive}. In addition, various types of problems in accelerator control and operation, such as fast and safe parameter tuning to achieve optimal beam energy in the SwissFEL~\cite{kirschner2019bayesian}, optics correction~\cite{fol2019optics} and collimator alignment in the LHC~\cite{azzopardi2019operational}, and model-free stabilization of source size in the ALS~\cite{leemann2019demonstration}, are intrinsically suitable for ML solutions due to the complexity of the input space, and a target function that is rather straightforward to implement. Applications in beam diagnostics have also attracted much interest due to ML's advantages on non-linear behavior and multi-objective optimization~\cite{fol2019jacow}.


Apart from the above-mentioned applications, anomaly detection and prevention has always been an inescapable and significant issue for accelerators to achieve high efficiency and reliability. There is also an increasing number of studies on data-driven approaches on this topic~\cite{tilaro2018model,piekarski2020convolutional}. However, most existing research focuses on diagnostic techniques that distinguish anomalies from normal circumstances, rather than prognostic approaches that predict potential anomalies in advance. Several recent studies have 
attempted to address the issue of prediction. Donon et al. present different approaches, including signal processing and statistical feature extraction, to identify and predict jitters in the RF sources of Linac4~\cite{donon2019anomaly,donon2020extended}. All approaches manage to detect the example jitters when they occur, and some even detect first symptoms ahead of time, which implies predictive power. Contrary to our beam interruptions, most jitters appear progressively rather than abruptly, which makes it more likely to locate an earlier precursor. There also lacks a quantified validation of the model performance among all the jitters. Rescic et al. applied various binary classification algorithms on beam pulses from the SNS to identify the last pulse ahead of the failure from normal pulses~\cite{rescic2020predicting}, in which predictive power is embedded, but their approach depends highly on the discrete pulse structure and cannot be directly adapted to continuous cases. It is also worth noting that both studies deal with univariate time series, i.e. RF power for Donon et al. and beam current for Rescic et al. Our work presents a novel and more ambitious approach to classify a continuous stream of multivariate time series data by cutting out windows from different regions, and trigger an alarm for a potential beam interruption before it happens. Model performance on the validation set is evaluated by two metrics, with examples from mimicked live prediction as well.

Outside the accelerator community, the prediction and prevention of failures has for long attracted researchers' interest, especially in the field of predictive maintenance~\cite{hashemian2010state, selcuk2017predictive} where a typical research deals with the lifetime prediction of an engine or bearing. 
However, in contrast to the gradual degradation process of a machinery equipment, the accelerator interruptions may appear rather suddenly due to momentary causes, thus traditional signal processing methods such as noise reduction or vibration analysis~\cite{scheffer2004practical,kovalev2018data}, are largely ineffective in the rare events scenario. Instead of long-term evolution in time, short-scale structures should raise more concern. By generating Recurrence Plots (RPs) in close adjacency of the failures as well as in the middle of stable operation period, we managed to extract finer information out of the original time series, which is better suited to the abrupt cases.

This work aims at capturing the precursor of beam interruptions - or for short "interlocks" - of the High Intensity Proton Accelerators (HIPA) by classifying whether the accelerator is in stable or unstable operation mode. A novel approach, a Recurrence Plots based Convolutional Neural Network (RPCNN), is introduced and adapted to the problem setting. The method transforms multivariate time series into images, which allows to extract enough structures and exploit the mature ML techniques of image classification. The RPCNN model achieves an AUC value of $0.71\pm 0.01$ and reduces $0.5\pm 0.2$ seconds of beam time loss per interlock, compared to an Area under the ROC Curve (AUC) value of $0.65\pm 0.01$ and $0.7\pm 0.1$ seconds of a Random Forest (RF) model. The RPCNN model predicts 4.9\% of interlock samples, and would potentially save 7.45 minutes more beam time during the HIPA run from September to December 2019. 

The paper is structured as follows: Section 2 introduces the dataset, the underlying algorithms and the evaluation metrics for presenting the result. Section 3 displays the results and interpretation. Finally Section 4 discusses the ambiguity and limitations as well as possible extensions of the work. 

%% file: 2.1_Dataset.tex
The HIPA at the Paul Scherrer Institut (PSI) is one of the most powerful proton cyclotron facilities in the world, with nearly 1.4 MW of beam power~\cite{reggiani2020improving}. The HIPA Archiver is an in-house developed archiver software that stores values of EPICS~\cite{dalesio1991epics} Process Variables, i.e. channels, according to tailored configuration settings~\cite{lutz2012database}. A data API is available as a Python library to access the archiver and export the data into Pandas Dataframes~\cite{ebner2020api}.

\begin{figure}[H]
\centering
\includegraphics[width=12cm]{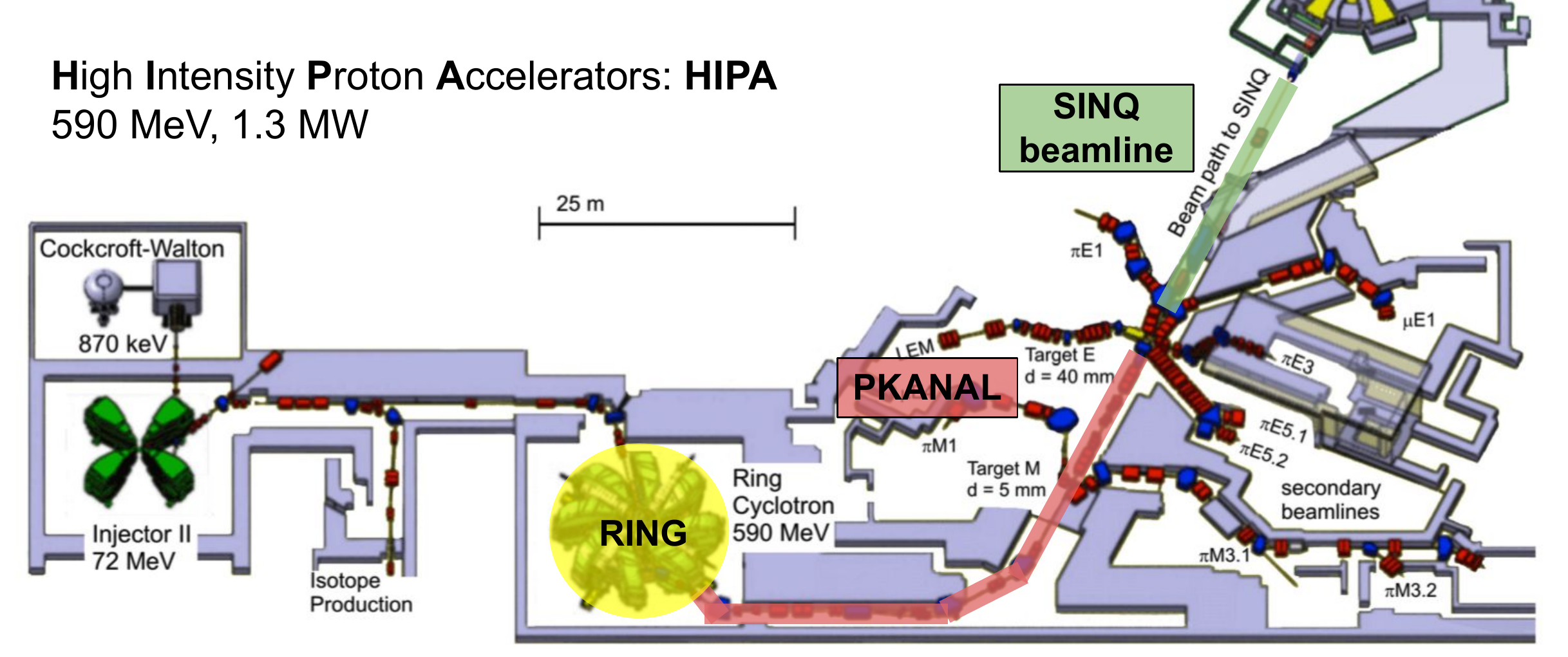}
\caption{Different HIPA sections of the channels.}
\label{fig:hipa_sections}
\end{figure}

\begin{figure}[H]
\centering
\includegraphics[width=15cm]{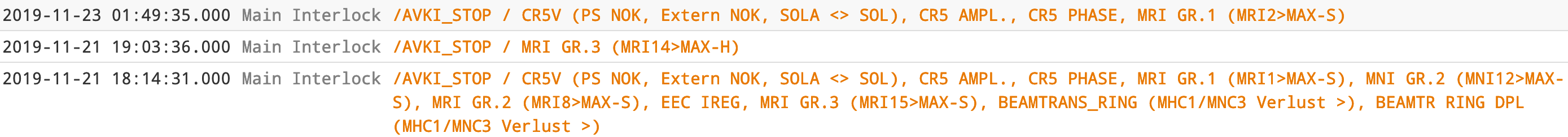}
\caption{Example interlock record messages.}
\label{fig:interlock_record}
\end{figure}

The dataset was taken from 14 weeks in 2019 starting from September from the HIPA Archiver, composed of 376 operating channels of the RING section and the PKANAL (see Figure~\ref{fig:hipa_sections}) section for the model input (SINQ beamline was not operating during the period), and the Interlock records for the model output. 
Figure~\ref{fig:hipa_sections}
shows the different sections of the channels, and table~\ref{tab:example_channels}
lists some example channels with detailed information. Figure~\ref{fig:interlock_record} shows some examples of typical interlock records, which are text messages reporting timestamp and all possible causes. Each interlock has at least one cause, though multiple causes can be attributed to one interlock if no single cause could be identified. For instance, an interlock of ``MRI GR.3'' indicates that at least one loss monitor at the end of Ring Cyclotron exceeds its limit, as shown in the second line of Figure~\ref{fig:interlock_record}. The detailed description ``MRI14>MAX-H'' gives the exact cause that the beam loss monitor ``MRI14'' is over its maximal allowed value. 

\begin{table}[H]
\caption{List of some example channels.}
\centering
\begin{tabular}{lll}
\toprule
\textbf{Channel} & \textbf{Section} &
\textbf{Description (unit)}\\
\midrule
AHA:IST:2 & PKANAL & Bending magnet ($\mathrm{A}$)  \\
CR1IN:IST:2 & RING & Voltage of Ring RF Cavity 1 ($\mathrm{kVp}$) \\
MHC1:IST:2 & PKANAL & Beam current measurement at the ring exit ($\mu \mathrm{A}$) \\
\bottomrule
\end{tabular}
\label{tab:example_channels}
\end{table}

\begin{figure}[H]
\centering
\includegraphics[width=7cm]{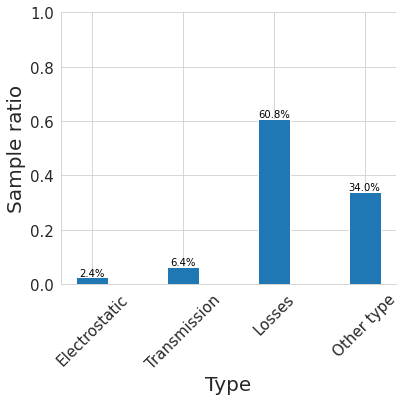}
\caption{Distribution of the interlock events by type. ``Other Types'' denotes all interlock types that are not prevalent enough to have a separate category. Note that an interlock may be labeled as more than one type. }
\label{fig:interlock_types}
\end{figure}

Taking into account the original causes of the interlocks, to ease the classification problem we separated them into four different general types.  Table~\ref{tab:interlock_types} and Figure~\ref{fig:interlock_types} show the four types and their statistics in the considered dataset that contains 2027 interlocks in total.
\begin{table}[H]
\caption{The different types of interlocks in the dataset.}
\centering
\begin{tabular}{ll}
\toprule
\textbf{Type} & \textbf{Description}  \\
\midrule
Electrostatic & Interlock related to electrostatic elements  \\
Transmission & Interlock related to transmission through target  \\
Losses & Interlock related to beam losses \\
Other type & Interlock related to another type or unknown\\
\bottomrule
\end{tabular}
\label{tab:interlock_types}
\end{table}

\begin{figure}[H]
\centering
\resizebox{12cm}{!}{%
\input{images/time_alignment}
 }
\caption{Synchronization of the beam current as an example channel and the interlocks.}
\label{figure:time_alignment}
\end{figure}
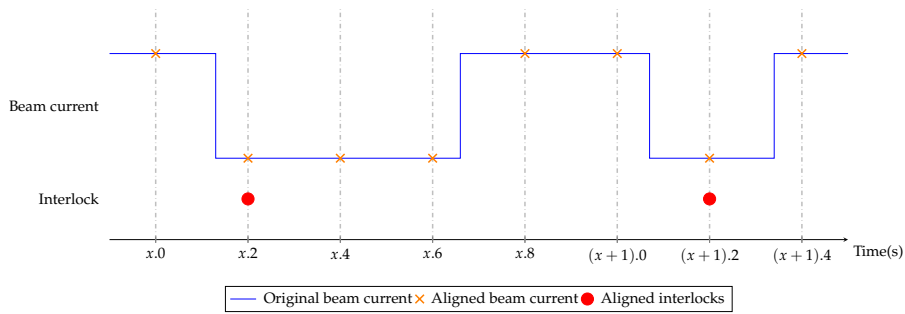

The archiver logs each value change with a certain maximum frequency per channel. In order to have input samples at a fixed frequency, during preprocessing the data of all channels are synchronized onto a predefined 5~Hz time grid by taking the last point that was recorded before each point in the grid (Figure~\ref{figure:time_alignment}).  The interlock timestamps as given by the archiver, were found to be up to a second off with respect to the real time of the beam interruption. Therefore, the interlock timestamp is chosen to be the closest point in the 5~Hz grid in which the beam current drops below a certain threshold. 


%% file: images/time_alignment.tex
\begin{tikzpicture}
    \begin{axis}[
        at={(0cm,0cm)},
        width=1.2\textwidth,
        height=0.45\textwidth,
        axis x line=center,
        axis y line=none,
        xlabel=Time(s),
        xmin=0, xmax=16, ymin=0, ymax=10,
        xtick={1,3,...,15},
        xticklabels={$x.0$,$x.2$,$x.4$,$x.6$,$x.8$,$(x+1).0$,$(x+1).2$,$(x+1).4$},
        xlabel style={at={(axis cs:16,0)}, anchor=north west},
        extra x ticks={1,3,...,15},
	    extra y ticks={0,5},
        extra x tick labels={},
        extra y tick labels={},
        extra tick style={grid=major, grid style={dash dot, line width=1pt}},
        legend columns=3,
        legend style={at={(axis cs:8,-2)},anchor=north,row sep=0.05cm,/tikz/every odd column/.append style={column sep=0.1cm}},
        legend cell align={left}
        ]
        \addplot[color=blue, forget plot]
        coordinates {(0,8)(2.3,8)(2.3,3.5)(7.6,3.5)(7.6,8)(11.7,8)(11.7,3.5)(14.4,3.5)(14.4,8)(16,8)};
        \addlegendimage{blue}
        \addlegendentry{Original beam current}
        \addplot[only marks, color=orange,mark=x,mark size=4pt,line width=1pt,forget plot]
        coordinates {
        (1,8)(3,3.5)(5,3.5)(7,3.5)(9,8)(11,8)(13,3.5)(15,8)
	    };
        \addlegendimage{only marks,color=orange,mark=x,mark size=4pt,line width=1pt}
        \addlegendentry{Aligned beam current}
        \node[anchor=east] (x) at (axis cs:0,5.75){};
        \addplot [only marks,color=red,mark size=4pt,forget plot] coordinates {
        (3, 1.75)
        (13, 1.75)
        };
        \addlegendimage{only marks,color=red,mark size=4pt}
        \addlegendentry{Aligned interlocks}
        \node[anchor=east] (int) at (axis cs:0,1.75){};
    \end{axis}
    \node[anchor=east] (x2) at (x){Beam current};
    \node[anchor=east] (int2) at (int){Interlock};
\end{tikzpicture}

%% file: 2.2_Problem.tex
The problem of interlock prediction is formulated as a classification problem of two types of samples -- \emph{interlock windows} labeled as class $1$ and \emph{stable windows} labeled as class $0$, taken from the channel time series. The model prediction of a sample is a value in $[0,1]$, and a sample is predicted to be class $1$ once its output gets above a custom classification threshold determined by performances.

\begin{figure}[H]
\centering
\includegraphics[width=12cm]{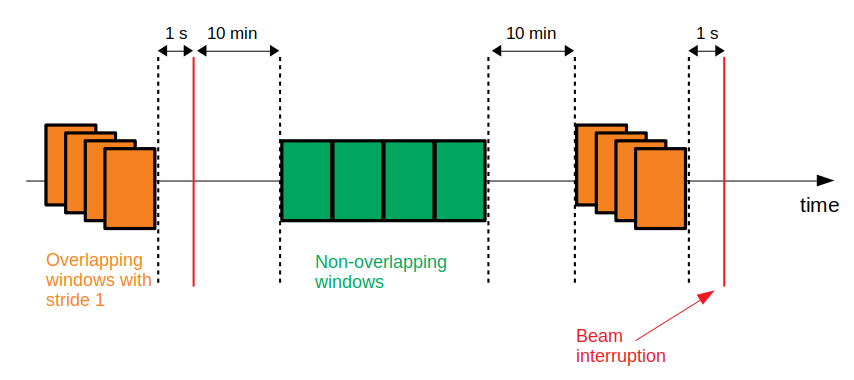}
\caption{Definition of interlock windows (orange) and stable windows (green). 5 overlapping interlock windows are cut at 1 second before the interlocks, and non-overlapping stable windows are cut in between two interlocks with a 10 minutes buffer region. Each window has a length of $12.8$ seconds ($64$ timesteps).}
\label{fig:window_cutting_method}
\end{figure}

An interlock window is taken by setting its endpoint closely before but not exactly at an interlock event. This allows the model to identify the precursors of the interlock and not the interlock event itself, in case the signals synchronization is slightly off. Thus, if the last timestamp of the window has been taken 1 second before the interlock event, the classification of a sample as interlock by the model means that an interlock is expected to happen in 1 second. To increase the number of interlock samples, an interval of 5 overlapping windows are assigned the label ``interlock'' rather than only one window per interlock, which is shown in the first two columns of Table~\ref{tab:number_of_samples}. 
The stable windows are taken as the series of non-overlapping windows that are in the periods between two interlocks. To stay away from the unstable situations, two buffer regions of 10 minutes before and after the interlocks are ignored, as displayed in Figure~ \ref{fig:window_cutting_method}.

Taking sliding windows as interlock samples mimics the real-time application scenario where there is a constant incoming data flow. Since the machine is mostly in stable operation mode, the stable samples are abundant and similar, thus it is neither necessary nor economical to cut sliding stable windows. All windows have a length of $12.8$ seconds ($64$ time steps, and each time step takes $0.2$ seconds), which is determined by the model performance.

In addition, data where the beam current (channel ``MHC1:IST:2'') presents a value of less than 1000~$\mu A$, at the first time stamp of the sample, are excluded from the dataset. This measure removes samples where the HIPA machine was not in a stable state. Such samples may not be a true representative of the respective class and thus not suitable as training data. 

As shown in Figure~\ref{fig:interlock_types}, interlocks of the type ``Losses'' make up a majority (60.8\%) of the available samples. Since the vast difference in types may be problematic for a binary classifier and as not enough samples of each type are available to perform a multi-class classification, only interlocks of the type ``Losses'' are considered in the current study. The number of considered interlocks reduces from 2027 to 894 after the above cleaning measures.

The model was trained on the first 80\% of the available data and validated on the remaining 20\% in time order to assess the predictive power of the model in future samples. 

On average, there is about one interlock event per hour in the dataset (the interlock distribution and counts during our data collection period are given in Appendix~\ref{apx:interlocks}). Due to the nature of the data, the two classes of windows are highly imbalanced, as shown in Table~\ref{tab:number_of_samples}. In order to compensate for the class imbalance, bootstrapping is employed on the training set. Interlock samples are drawn with replacement from the training set until their number equals the number of stable samples. Numbers of interlock and stable samples in training and validation sets after all preprocessing steps with and w/o bootstrapping are listed in Table~\ref{tab:number_of_samples}.

\begin{table}[H]
\caption{Numbers of samples in the training and validation sets, before and after bootstrapping.}
\centering
\begin{tabular}{lllll}
\toprule
     & \textbf{\# interlock events}	& \textbf{\# interlock samples} & \textbf{\# stable samples} &
     \begin{tabular}[x]{@{}c@{}}\textbf{\# interlock samples} \\ \textbf{after bootstrapping}\end{tabular} \\
\midrule
\textbf{Training set} & $731$ & $3655$ & $176046$& $176046$ \\
\textbf{Validation set}&  $163$ & $815$ & $44110$ & not applied \\
\bottomrule
\end{tabular}
\label{tab:number_of_samples}
\end{table}

%% file: 2.3.1_Recurrence_theory.tex
\begin{figure}[H]
    \centering
    \includegraphics[width=12cm]{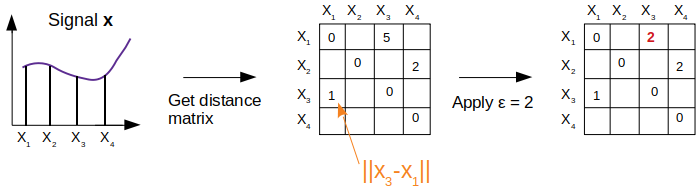}
    \caption{Generation of recurrence plot from a signal with fixed $\epsilon=2$.}
    \label{fig:rp_explanation}
\end{figure}

Recurrence plots were developed as a method to analyze dynamical systems and detect hidden dynamical patterns and nonlinearities \cite{eckmann1987recurrence}. The structures in a recurrence plot contain information about time evolution of the system it portrays. In this study, recurrence plots are used to transform the time series into images, which are then classified by a Convolutional Neural Network. 

The original recurrence plot described in \cite{eckmann1987recurrence} is defined as
\begin{equation*}
	R_{i,j} = \theta(\epsilon_i - ||\vec{x}_i-\vec{x}_j||),\hspace{5pt} \vec{x}_i\in \mathbb{R}^m, i,j=1,\dots,N.
\end{equation*}
with $\theta$ the Heaviside function, $i,j$ the indices of time steps inside a time window taken from the signal, and $N$ the length of the window. Here the radius $\epsilon_i$ is chosen for each $i$ in such a way that the neighborhood it defines contains a fixed number of states $\vec{x}_j$. As a consequence the recurrence plot is not symmetric but all columns have the same recurrence density. The most common definitions use a formulation with a fixed radius $\epsilon_i = \epsilon, \hspace{5pt} \forall i$ which was first introduced by Zbilut et al. in ~\cite{zbilut1990use}. The variation we use here is a so-called global recurrence plot~\cite{webber2005recurrence} with a fixed epsilon, as defined in Equation~\eqref{eq:distance_matrix}
\begin{equation}
	D_{i,j}= \begin{cases}
	||\vec{x}_i - \vec{x}_j||, \quad ||\vec{x}_i - \vec{x}_j|| \leq \epsilon \\
	\epsilon,\quad\quad\quad\quad\quad ||\vec{x}_i - \vec{x}_j|| > \epsilon	
		\end{cases}
		\label{eq:distance_matrix}
\end{equation}
where $D$ is symmetric. Figure~\ref{fig:rp_explanation} explains the process of transforming an original signal to a recurrence plot, with fixed $\epsilon=2$.

\begin{figure}[H]
    \centering
    \includegraphics[width=15cm]{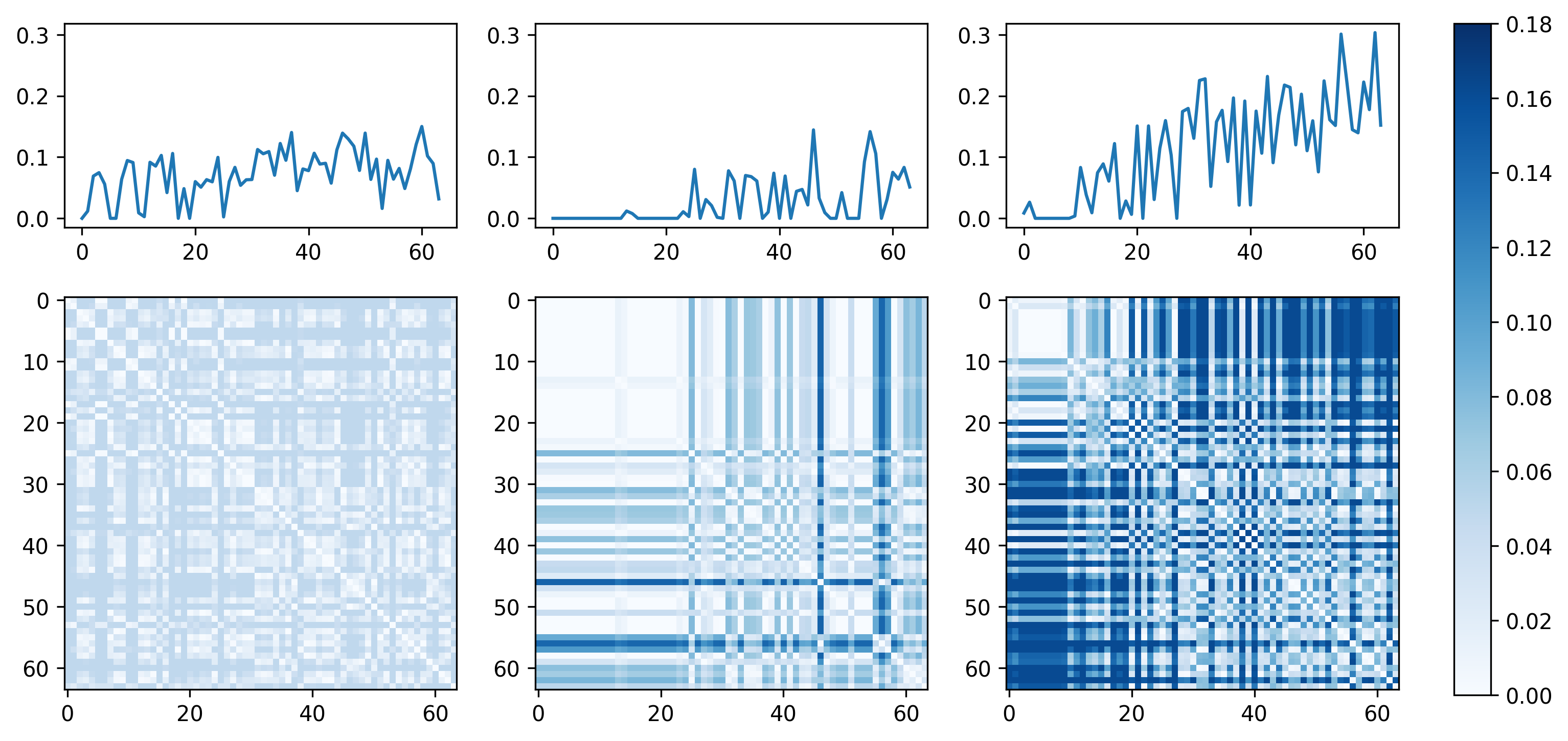}
    \caption{Examples of recurrence plots generated from the RPCNN model. From left to right: uncorrelated stochastic data, data starting to grow, and stochastic data with a linear trend. The top row shows the signals before the recurrence plot generation, and the bottom row shows the corresponding recurrence plots.}
    \label{fig:rp_example}
\end{figure}

The patterns of recurrence plots convey a wealth of information not directly available from the time series they are based on, such as white areas or band structures in Figure~\ref{fig:rp_example}, which indicates abrupt changes~\cite{website:Recurrence_Plot, marwan2007recurrence}. Typical methods to extract such information include Recurrence Quantification Analysis (RQA)~\cite{webber2015recurrence}, which deals with the analysis of the small-scale structures in recurrence plots. The RQA presents a number of metrics to quantify the recurrences, such as the recurrence rate, trapping time or divergence, furthermore RQA can be applied to non-stationary or very short time-series as are present in this study. Instead, we choose to feed the recurrence plots into a CNN on account of its great success on image classification. CNNs are capable of constructing novel and optimal features beyond the limits of RQA's predefined metrics. 

%% file: 2.3.2_Recurrence_methods.tex

The RPCNN model is constructed in TensorFlow using the Keras API~\cite{chollet2015keras}. The architecture of the RPCNN model is outlined in Figure~\ref{fig:model_architecture}. Details are presented in Appendix~\ref{apx:rpcnn}. Table \ref{tab:RPCN_parameters} lists the training parameters and a selection of the layer settings. 

\begin{figure}[H]
    \centering
    \includegraphics[width=12cm]{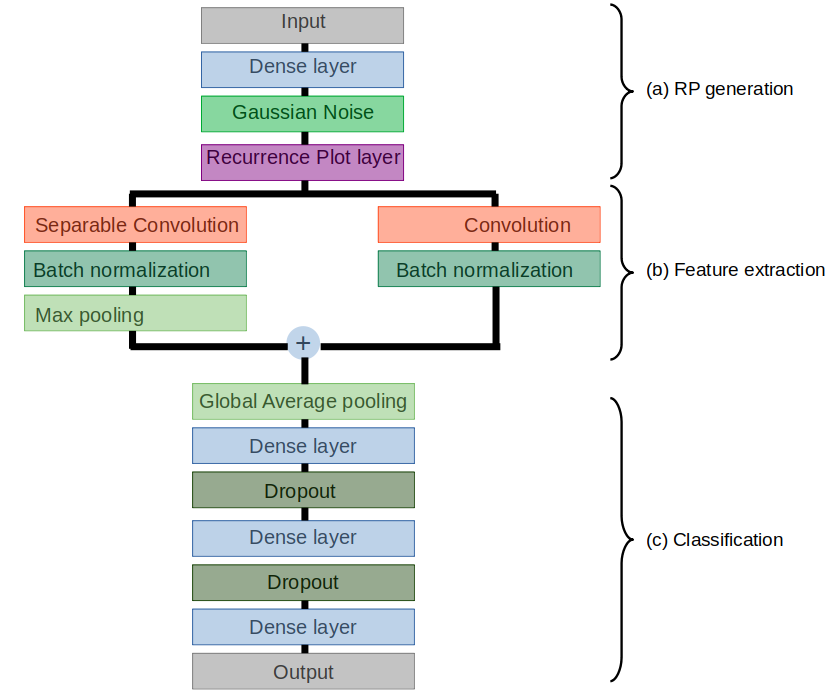}
\caption{Schematic representation of the RPCNN model architecture.}
\label{fig:model_architecture}
\end{figure} 

\begin{table}[H]
\caption{Overview of the RPCNN model training parameters.}
\centering
\begin{tabular}{ll}
\toprule
\textbf{Parameter} & \textbf{Value}\\
\midrule
Learning rate & 0.001 \\
Batch size & 512 \\
Optimizer & Adam \\ 
Window size & $12.8~s$ \\
Number of channels & $97$\\
Std of the Gaussian noise & $0.3$ \\
Dropout ratio & $0.1$ \\
\bottomrule
\end{tabular}
\label{tab:RPCN_parameters}
\end{table}

\paragraph{Preprocessing}
In order to decrease the model parameters and thus reduce over-fitting, only $97$ of the available $376$ channels were used as model input. Since there are groups of input channels that measure approximately the same quantity and have a high correlation between each other (for instance beam position monitors),
a lot of redundancy exists in the input channels, which enables an automated random feature selection procedure based on model performance. The selection was done by a combination of random search and evaluation of past experimental results, with details presented in Algorithm~\ref{alg:selection}. Note that the resulting set of channels is only one out of many possible channel combinations.

\begin{algorithm}[H]
\SetAlgoLined
N = total number of channels,
n = sampled number of channels,
M = total number of runs\;
Fix the initialization by fixing the Tensorflow seed\;
\For{n in a list of randomly chosen numbers < N}{
    \For{M runs}{randomly sample n out of N channels\;
    train a model with the sampled channels\;
    calculate beam time saved for that model\;
    }
   }
Take the set of channels which optimize the beam time saved, considering model convergence and losses\;
Repeat the subsampling among the currently selected set, to obtain the best performing set\;
Randomly change the initialization seed with the channels fixed to the best performing set, to obtain the best performing configuration\;
\caption{Random feature selection}
 \label{alg:selection} 
\end{algorithm}

The channel signals that are used as input features differ vastly with respect to their scale. As this might impact the convolution operations, the channel signals are standardized to mean $0$ and standard deviation $1$. 

\begin{figure}[H]
    \centering
    \includegraphics[width=7cm]{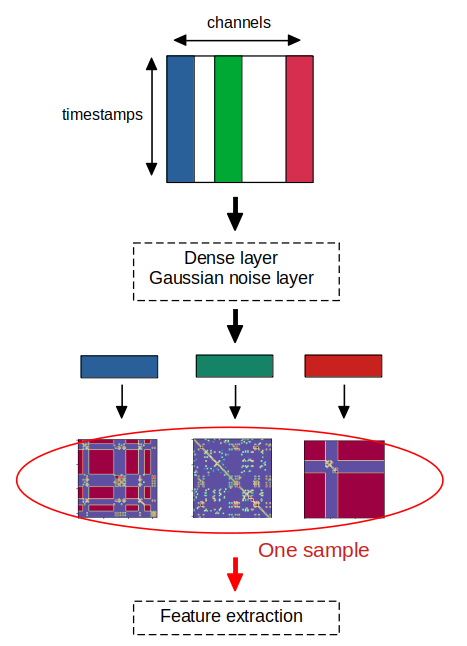}
\caption{Illustration of section \textbf{(a) RP generation} of the RPCNN model (c.f. Figure~\ref{fig:model_architecture}).}
\label{fig:sample_illustration}
\end{figure}

\paragraph{RP generation}
The RP generation part starts with a $L_{2}$-regularized fully connected layer to reduce the number of available features from $97$ to $20$. Gaussian noise with a standard deviation of $0.3$ is added as a regularization and data augmentation measure before the signals are converted into recurrence plots. The process is illustrated in Figure~\ref{fig:sample_illustration}.

The recurrence plots are produced internally by a custom ``Recurrence Plot layer'' in the model. This approach allows to set trainable $\epsilon$ parameters and generate recurrence plots on the fly, which eases deployment of the trained models as the recurrence plots do not have to be generated and stored explicitly during preprocessing. For each feature a recurrence plot is drawn and the respective $\epsilon$ parameters are initialized with random numbers in the interval $[0,1]$. Examples of recurrence plots of an interlock sample and a stable sample are given in Appendix~\ref{apx:recurrence}, where each sample contains $20$ plots respectively. 

\paragraph{Feature extraction}
The structure of the feature extraction section was adapted from~\cite{chollet2017xception} and consists of a depthwise separable convolution performed in parallel with a pointwise convolution. Both are followed by a ReLU non-linearity. Batch normalization layers are added after both convolutional layers.

\paragraph{Classification}
The classification part of the network consists of three fully connected layers separated by dropout layers with a ratio of $0.1$. $L_2$ regularization was applied to all fully connected layers.

%% file: 2.4_RF.tex
As a feasibility study of the dataset and baseline model for the RPCNN, we trained another Random Forest~\cite{ho1995random} model on all 376 input channels.
The Random Forest classifier is an ensemble learning method built upon Decision Trees. While each tree decides its optimal splits based on a random sub-sample of training data and features, the ensemble takes the average classification scores from all trees as the output classification. It is widely applied due to its robustness against over-fitting, straightforward implementation and relatively simple hyper-parameter tuning. It also does not require the input features to be standardized or re-scaled, and no difference is noticed on trial runs.

For the RPCNN model, information on time dimension is embedded in the recurrence plots implicitly. It is thus incompatible to use a window with both time and channel dimensions directly as input for the RF model. Instead, only the last timesteps of each of the same interlocks and stable windows are fed as input into the RF model. The RF model is implemented using the RandomForestClassifier method of the Scikit-learn~\cite{scikit-learn} ensemble module with parameters listed in Table~\ref{tab:RF_parameters}. The RF on the 97 channels from the RPCNN shows the same performance as with all 376 channels.

\begin{table}[H]
\caption{Overview of the RF model training parameters.}
\centering
\begin{tabular}{ll}
\toprule
\textbf{Parameter} & \textbf{Value}\\
\midrule
Number of estimators & 90 \\
Maximum depth & 9 \\ 
Number of channels & $376$\\
Maximum leaf nodes & $70$ \\
Criterion & Gini \\
\bottomrule
\end{tabular}
\label{tab:RF_parameters}
\end{table}

%% file: 2.5_Metric.tex
Typical binary classification metrics in a confusion matrix are defined and applied in our setting. A True Positive (TP) means an interlock sample -- a sample less than 15 seconds before an interlock -- being classified as an interlock. A False Positive (FP) means a stable sample -- a sample at least 10 minutes away from an interlock -- being mistaken as an interlock. A True Negative (TN) means a stable sample being reported as stable. And the rest False Negative (FN) means that the model fails to identify an interlock. All the metrics are counted with regard to samples, namely $TP+FN=815$ for validation set according to Table~\ref{tab:number_of_samples}. 

For an imbalanced dataset such as the present one, evaluation of the model performance through the classification accuracy ($\frac{TP+TN}{TP+FN+FP+TN}$ ) is not sufficient. The performance of the model is evaluated with a Receiver Operating Curve (ROC) as well as a target defined to maximize the beam time saved. 

\subsubsection{Receiver Operating Curve}
The ROC curve shows the true positive rate ($TPR=\frac{TP}{TP+FN}$) against the false positive rate ($FPR=\frac{FP}{FP+TN}$) of the predictions of a binary classification model as a function of a varying classification threshold~\cite{fawcett2006introduction}.
The AUC measures the area under the ROC curve bounded by the axes. It represents the probability that a random positive (i.e. interlock) sample receives a higher value than a random negative (i.e. stable) sample, which ranges from $0$ to $1$ and $0.5$ corresponds to random guess. Aggregated over all possible classification thresholds, it indicates the general capacity of a model to distinguish between the two classes.

\subsubsection{Beam time saved}
We also propose a more practical metric for this particular use case. The beam is shut off when an interlock occurs. After each interlock the beam current is ramped up automatically again. Each interlock event causes a beam time loss of about 25~seconds. The average uptime of the HIPA facility is about 90\%~\cite{HIPAoperation}. Short interruptions, mostly from beam losses and electrostatic elements, are responsible for about 2\% beam time lost at HIPA. Hence an ``actually negative'' state (i.e. $FP+TN$) is about 45 times (90\% over 2\%)  more likely to occur than an ``actually positive'' state (i.e. $TP+FN$). We denote $p$ as the probability of ``actually positive'', i.e. the probability that an interlock would occur. Thus $p = \frac{2\%}{2\%+90\%}=\frac{1}{46}$.

For the target definition it is assumed that each correctly predicted interlock can be prevented by a 10\% beam current reduction for 60~seconds. Each interlock thus leads to an equivalent of six seconds of lost beam time. Note that this is only for performance evaluation purposes and the 10\% reduction is not necessarily able to prevent interlocks in real life. The actual mitigation measure is still under investigation.

\begin{figure}[H]
    \centering
    \includegraphics[width=12cm]{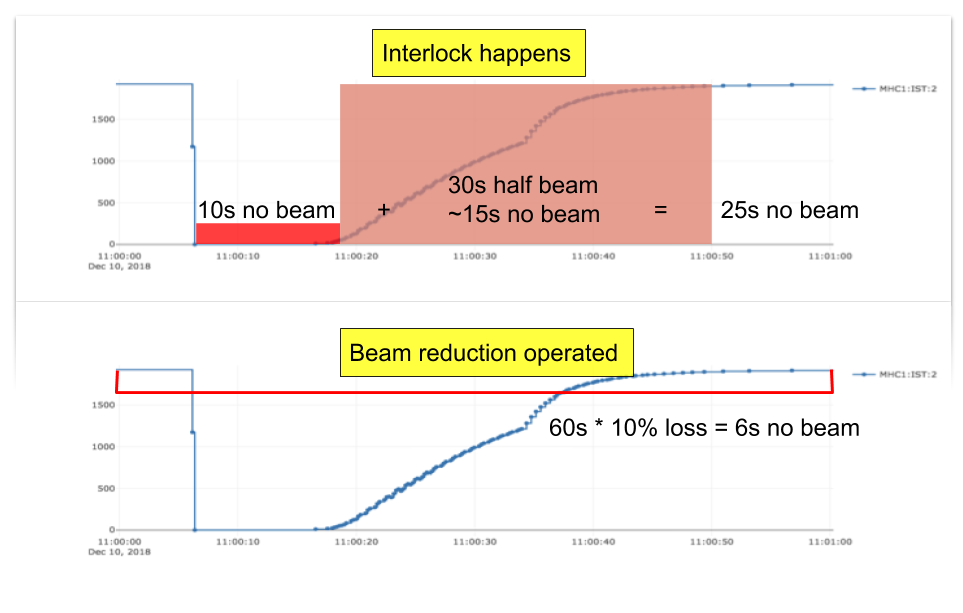}
    \caption{Expected lost beam time without (top) and with (bottom) the proposed 10\% current reduction. An interlock equivalently leads to 25 seconds of lost beam time. With the current reduction, the interlock is expected to be avoided with a cost of six seconds of lost beam time.}
\label{fig:current_reduction}
\end{figure}

Under this assumption, the evaluation of beam time saved is illustrated in Figure~\ref{fig:current_reduction} and shown in detail in Table~\ref{tab:beam_time_saved}.

\begin{table}[H]
\caption{Detail of beam time saved according to different classification result of one sample.}
\centering
\begin{tabular}{lllll}
\toprule
\textbf{$T_s$ (s)} & \textbf{TP} & \textbf{FN} & \textbf{FP} & \textbf{TN}\\
\midrule
Without current reduction & -25 & -25 & 0 & 0\\
With current reduction & -6 & -25 & -6 & 0 \\ 
Incentive & 19 & 0 & -6 & 0\\
\bottomrule
\end{tabular}
\label{tab:beam_time_saved}
\end{table}

Based on the above analysis, the ``beam time saved'', denoted by $T_s$, per interlock is defined in Equation~\eqref{eq:beam_time_saved}. We decide on the best classification threshold corresponding to the largest $T_s$, rather than the largest AUC value.
\begin{align}
    T_s &= 19 \cdot \frac{TP}{\# \text{ interlocks}} - 6 \cdot \frac{FP}{\# \text{ interlocks}} \nonumber \\
    &= 19 \cdot \frac{TP}{TP + FN} - 6 \cdot \frac{FP}{TP + FN} \nonumber \\
    &=19 \cdot TPR - 6 \cdot FPR \cdot \frac{FP + TN}{TP + FN} \nonumber \\
    &= 19 \cdot TPR  -  6 \cdot FPR \cdot \frac{1-p}{p} \nonumber \\
    &= 19 \cdot TPR - 270 \cdot FPR \quad \text{measured in }(s/\text{interlock})
\label{eq:beam_time_saved}   
\end{align}

%% file: 3.1_Results_evaluation.tex
\begin{figure}[H]
    \centering
    \begin{minipage}[c]{0.49\linewidth}
        \centering
        \includegraphics[width=\linewidth]{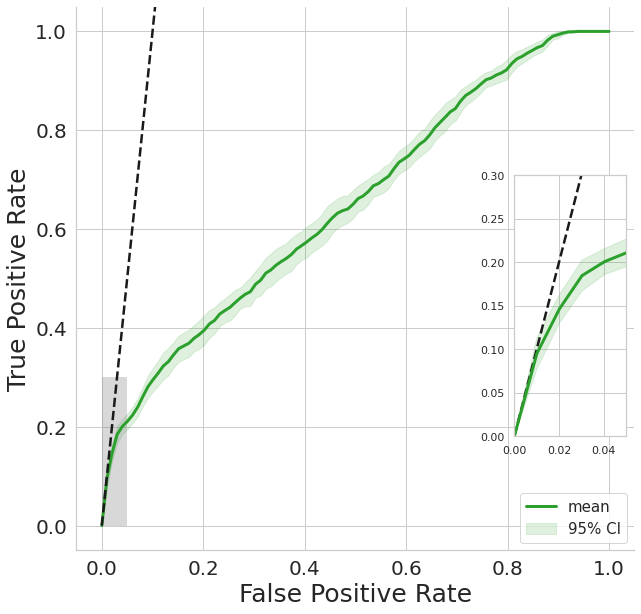}\\
        (a) RF best
    \end{minipage}
    \begin{minipage}[c]{0.49\linewidth}
        \centering
        \includegraphics[width=\linewidth]{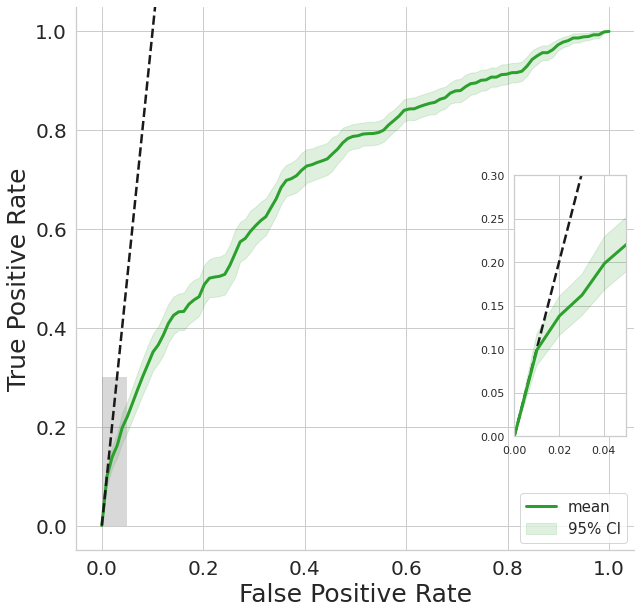}\\
        (b) RPCNN best
    \end{minipage}
    \begin{minipage}[c]{0.49\linewidth}
        \centering
        \includegraphics[width=\linewidth]{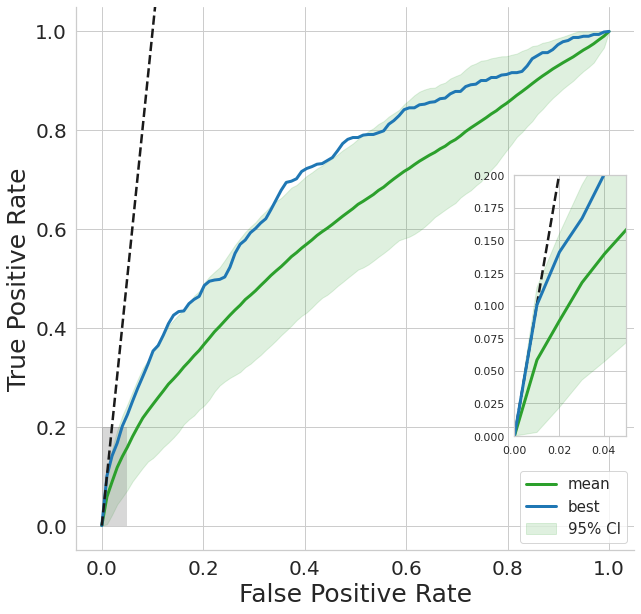}\\
        (c) RPCNN varying initialization
    \end{minipage}
\caption{ROC curves of the best-performing models in terms of Beam time saved and AUC: (\textbf{a}) RF; (\textbf{b}) RPCNN. The mean and 95\% confidence interval is calculated from re-sampling of the validation sets. (\textbf{c}) Varying initialization of the same RPCNN model as in (\textbf{b}). The mean and 95\% confidence interval are calculated from the validation results of 25 model instances.} 
\label{fig:roc_models}
\end{figure}

Figure~\ref{fig:roc_models} shows the ROC curves of the best performing RF and RPCNN models, as well as the ROC curve over varying initializations of the RPCNN model. 

The confidence intervals are a measure for the variability of the model, which may result from different selection of validation dataset, or randomness in model and parameter initialization. The variability over configurations of the validation dataset was calculated for both the RF and RPCNN model in Figure~\ref{fig:roc_models} \textbf{(a)} and \textbf{(b)}. Specifically, the models are trained on the same training set, then validated 20 times over randomly sampled subsets of the validation set. Initialization in the RPCNN model could also introduce various uncertainties depending on the convergence behavior of the model. Therefore, a corresponding confidence interval of the RPCNN model is calculated. For Figure~\ref{fig:roc_models} \textbf{(c)} 25 RPCNN models were trained and evaluated using identical training and validation sets. The mean results with the confidence intervals, as well as the same best-performing model shown in Figure~\ref{fig:roc_models} \textbf{(b)} are displayed.

The black dashed line marks the boundary $19\cdot TPR-270\cdot FPR=0$. A model could only save beam time if its ROC curve reaches the left side of this dashed line and saves more beam time the further away it is from this line.

\begin{table}[H]
\caption{Beam time saved ($T_s$) and AUC results of the best-performing RF and RPCNN models over re-sampling of the validation sets.}
\centering
\begin{tabular}{lll} 
\toprule
\textbf{Model} & \textbf{$T_s$ [s/interlock]}	& \textbf{AUC} \\
\midrule
RF best& $\mathbf{0.7\pm 0.1}$ & $0.65\pm 0.01$ \\ 
RPCNN best & $0.5\pm 0.2$ & $\mathbf{0.71\pm 0.01}$\\ 
RPCNN mean over initialization & $0.1\pm 0.1$ & $0.61\pm 0.04$ \\
\bottomrule
\end{tabular}
\label{tab:result_model} 
\end{table} 

\begin{table}[H]
\caption{Classification results in terms of True Positive (TP), False Positive (FP) and False Negative (FN) of the best-performing RF and RPCNN models. The thresholds for both models were obtained from the ROC curves, with $0.70$ for the RF and $0.65$ for the RPCNN.}
\centering
\begin{tabular}{llll} 
\toprule
\textbf{Model} &  \textbf{TP}& \textbf{FP}	& \textbf{FN} \\
\midrule
RF best & $25$ & $28$ & $790$ \\
RPCNN best & $40$ & $75$ & $775$\\ 
\bottomrule
\end{tabular}
\label{tab:confusion_matrix} 
\end{table} 

The AUC and beam time saved for the models displayed in Figure~\ref{fig:roc_models} can be found in Table~\ref{tab:result_model}. The TP, FP and FN counts for the best performing RF an RPCNN models are shown in Table~\ref{tab:confusion_matrix}. The RF model saves $0.7$ s per interlock, better than $0.5$ s of the RPCNN, but the RPCNN achieves a higher AUC value. 
The RPCNN successfully predicts $4.9\%$ of interlock samples ($40$ out of $815$). While it correctly identifies more interlocks than the RF ($40$ vs. $25$), it also triggers more false alarms ($75$ vs. $28$).

During the HIPA run from September to December 2019, there are $894$ interlocks of the ``Losses'' type, ignoring the secondary ones. Taking the result of the best RPCNN model, $0.5$ seconds would be saved for each interlock, which means we would deliver $7.45$ minutes more beam altogether. 
 

%% file: 3.2_Results_live.tex
\begin{figure}[H]
    \centering
    \begin{minipage}{0.48\linewidth}
        \centering
        \includegraphics[width=\linewidth]{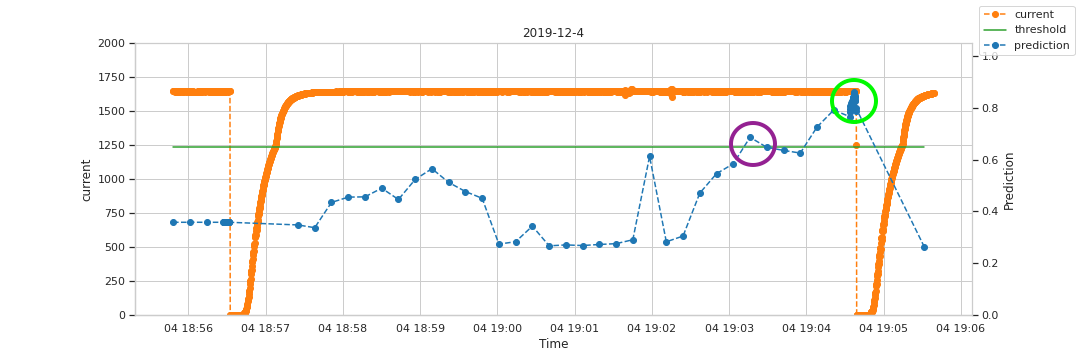}\\
    \end{minipage}
    \begin{minipage}{0.48\linewidth}
        \centering
        \includegraphics[width=\linewidth]{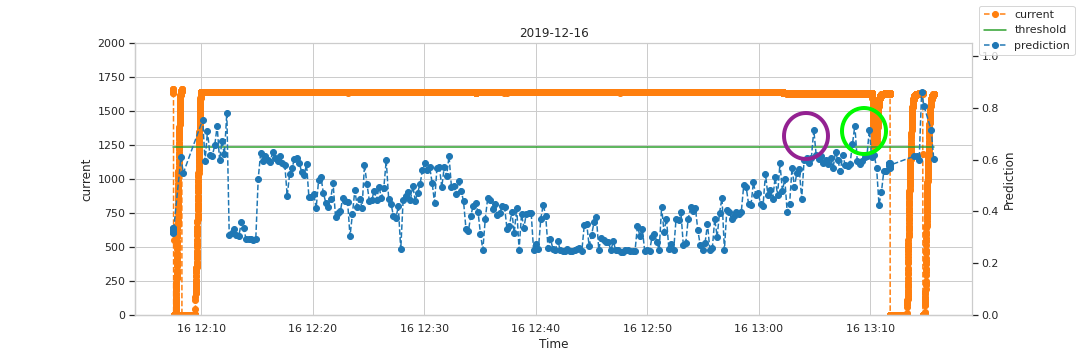}\\
    \end{minipage}
    \begin{minipage}{0.48\linewidth}
        \centering
        \includegraphics[width=\linewidth]{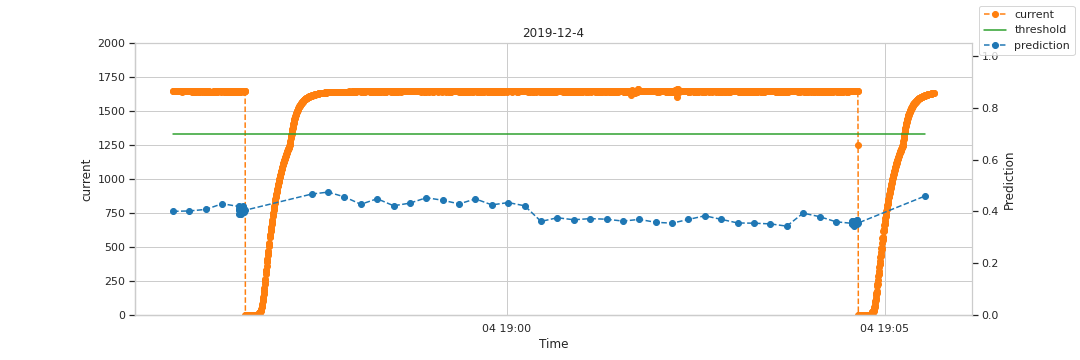}\\
        (a) TP from RPCNN (top) \\ compared to RF (bottom)
    \end{minipage}
    \begin{minipage}{0.48\linewidth}
        \centering
        \includegraphics[width=\linewidth]{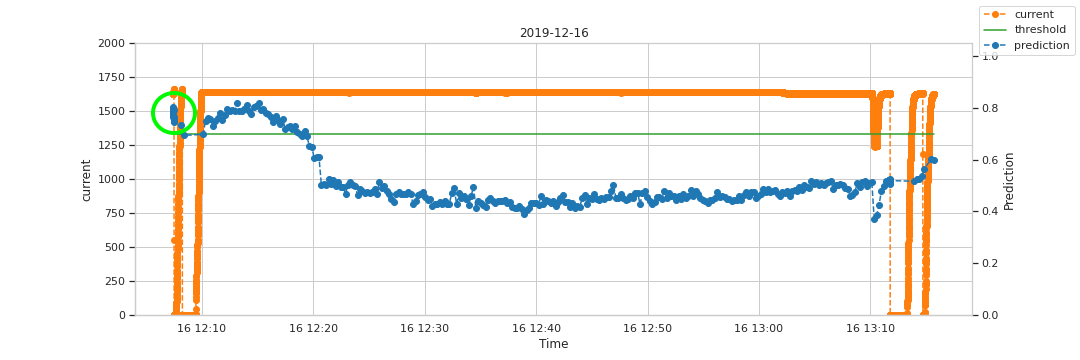}\\
        (b) TP from RPCNN (top) \\ compared to RF (bottom)
    \end{minipage}
    \begin{minipage}{0.49\linewidth}
        \centering
        \includegraphics[width=\linewidth]{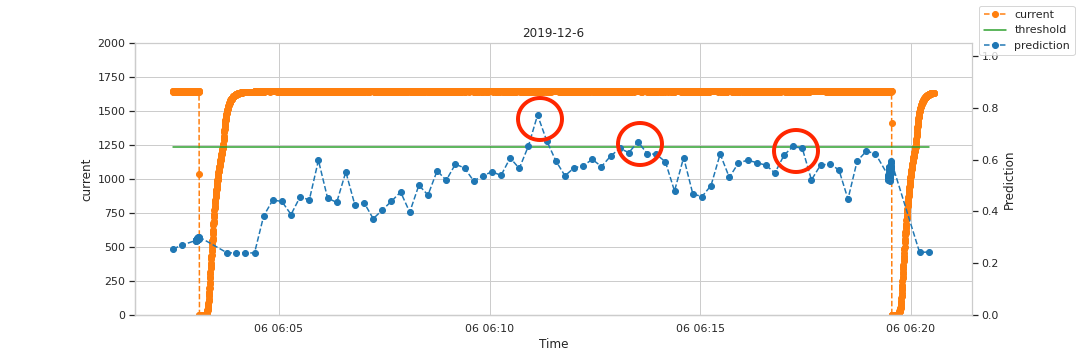}\\
    \end{minipage}
    \begin{minipage}{0.49\linewidth}
        \centering
        \includegraphics[width=\linewidth]{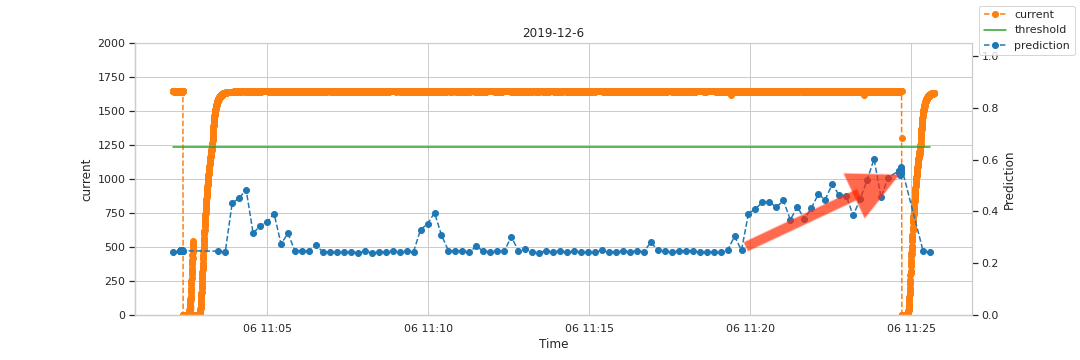}\\
    \end{minipage}
    \begin{minipage}{0.49\linewidth}
        \centering
        \includegraphics[width=\linewidth]{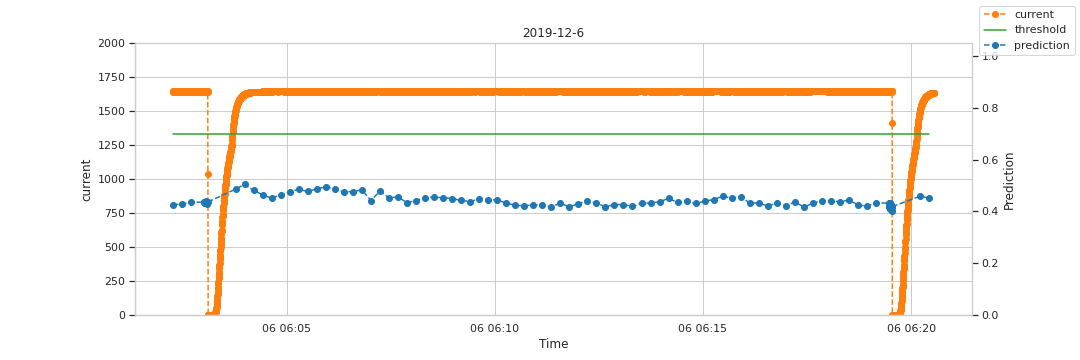}\\
        (c) FP from RPCNN (top) \\ compared to RF (bottom)
    \end{minipage}
    \begin{minipage}{0.49\linewidth}
        \centering
        \includegraphics[width=\linewidth]{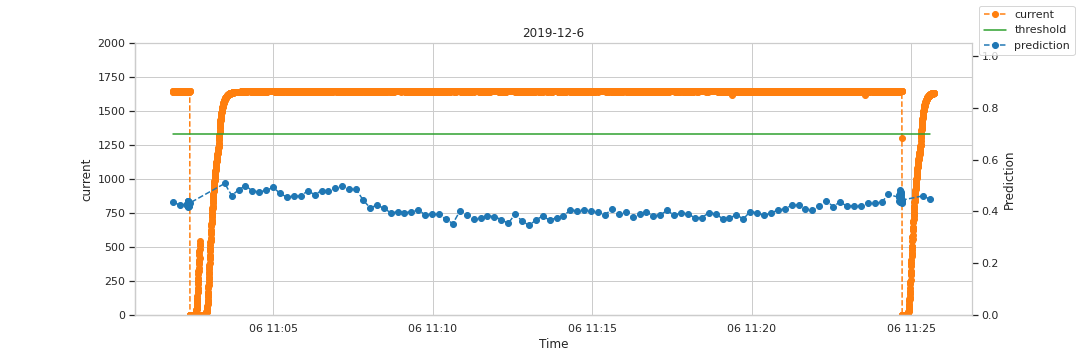}\\
        (d) FN from RPCNN (top) \\ compared to RF (bottom)
    \end{minipage}
\caption{Screenshots from simulated live predictions over the validation data. The blue line is the prediction value from the model, with higher values indicating higher probability that interlocks are approaching. The orange line is the readout of the beam current channel ``MHC1:IST:2'' where a drop to zero indicates an interlock. The green line is the binary classification threshold, here taking $0.65$ for the RPCNN and $0.7$ for the RF. (\textbf{a}) and (\textbf{b}) show two examples of successful prediction of interlocks by RPCNN compared to the result of RF on the same time periods. The positive predictions closer to interlocks are regarded as TP marked by green circles, and the earliest times that the model output cross the threshold are enclosed in purple circles. (\textbf{c}) shows the FPs from the RPCNN in red circles, compared to the result of RF. (\textbf{d}) shows an example of FN from the RPCNN compared to the result of RF. A clear trend increasing with time is present in the output of RPCNN, shown by the red arrow.}
\label{fig:tpfp_examples}
\end{figure}

While the AUC and Beam time loss metrics allow an evaluation of the expected performance of the model, the final test is the behavior of the model during live, i.e. continuous predictions. Figure~\ref{fig:tpfp_examples} presents a selection of snapshots of the RPCNN and RF models during mimicked live predictions on the extended validation set -- to examine the model behavior in unseen regions, the buffer regions are reduced from $10~\si{min}$ to $14~\si{s}$ and predictions are generated until $0.2~\si{s}$ before the interlock event. 

Figure~\ref{fig:tpfp_examples} shows that the RPCNN model achieves an overall better performance than the RF model in this mimicked live setting. In the successful cases of \textbf{(a)} and \textbf{(b)}, the RPCNN not only reports imminent predictions of interlocks in time scale of seconds (green circle), as how the training set is defined, but could also generate alarms farther in advance (purple circle), even several minutes ahead of time. However, there is much subtlety in the distinction between an early alarm and a FP according to the time from the earliest threshold crossing until the interlock. The three crossings (red circle) in \textbf{(c)} could be either taken as FPs if the machine is believed to remain stable and only start deviating shortly before interlocks, or recognized as early precursors if the machine keeps operating in an unstable mode. As for the prediction failure in \textbf{(d)}, the RPCNN still outperforms RF in the sense of a growing trend in output values (red arrow), which indicates some underlying progress of anomalies.

\begin{figure}[H]
    \centering
    \includegraphics[width=9cm]{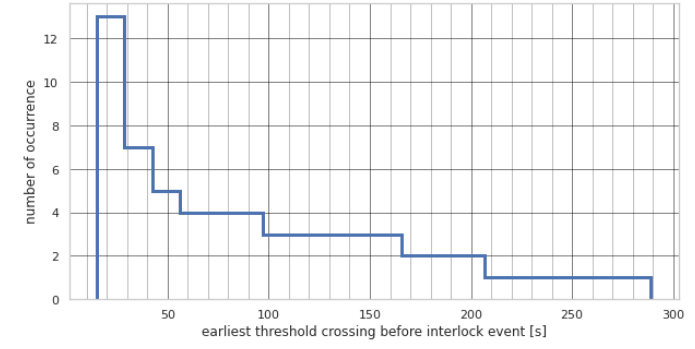}
\caption{Reverse cumulative histogram of the first instance the alarm threshold was crossed before an interlock event and the model prediction values remained above the threshold. From evaluation of the ROC curve a threshold value of $0.65$ is chosen. Displayed are the results for the best performing RPCNN model. Note that the interlock event is at 0~s, i.e time runs from right to left.}
\label{fig:cummulative_histogram_RPCNN} 
\end{figure} 

The behavior of the RPCNN model as seen in Figure~\ref{fig:tpfp_examples} suggests that some interlocks are detected well before the $1~s$ range defined in the problem formulation. This observation is supported by the findings shown in Figure~\ref{fig:cummulative_histogram_RPCNN}. The time range before the interlock event at which the RPCNN model prediction first crossed and stayed above the threshold is obtained for interlocks in the validation set and compiled into a reverse cumulative histogram. Detection times vary between the interlock events with some events being only detected a second before their occurrence and others close to 5~min beforehand.

%% file: 4_Discussion.tex
\unskip
\subsection{Achievements}
As shown in Table~\ref{tab:result_model}, all variations of models achieve an AUC value larger than $0.6$, indicating that our models possess the ability to distinguish between potential interlocks and stable operation. The customized target $T_s$, the ``beam time saved'', poses strict requirements on the model performance since it has little tolerance for FPs, as described in Equation~\eqref{eq:beam_time_saved}. In spite of such constraints, our models is able to increase the amount of
beam for the experiments. 

It is observed from Figure~\ref{fig:tpfp_examples} that the RPCNN model, while showing comparable performance with the RF on binary classification metrics in Table~\ref{tab:result_model}, is more suitable for real-time predictions.

\subsection{Limitations}
Despite the mentioned achievements and potential, there are still limitations in multiple aspects. As shown in Figure~\ref{fig:roc_models} \textbf{(c)}, the RPCNN model under random initialization does not converge well, which is expected to be resolved by acquisition of more taining data.  Moreover, the available channels might change for each data run. For example the 2019 model is invalid in 2020, since the SINQ beamline was restarted, so that several old channels were no longer available. 

Another limitation lies in the ambiguity in manipulating the data and model to achieve better training, from sample selection (number and size of windows, buffer region etc.), extent of preprocessing and data augmentation, feature engineering considering the redundancy in channels, to the selection of model architecture and hyper-parameter values.

Although a thorough cross-validation is a more convincing proof of the model performance, the confidence interval in Figure~\ref{fig:roc_models} \textbf{(a)} and  \textbf{(b)} are computed solely from constructing different subsets of the validation dataset. A cross-validation is done for the RF model with same level performance. For the RPCNN model, there has not yet been a complete cross-validation over the full dataset, since the model still diverges from initialization. Future studies should address these issues with new data.

\subsection{Outlook}

The results obtained in a mimicked live setting as displayed in Figure~\ref{fig:tpfp_examples} and Figure~\ref{fig:cummulative_histogram_RPCNN} indicate that the problem formulation as well as the performance metrics need to be reevaluated. As the detection times of the interlock events are inconsistent, labeling a fixed number of 5 samples per event as interlock might not be optimal. A next step could be a closer investigation to determine if the detection time can be linked to some property of the interlocks like their type, length or location, and to adjust the sample labeling accordingly.

Following the discussion in~\ref{sec:live} about the FPs in Figure~\ref{fig:tpfp_examples} \textbf{(c)}, another adaptation could be changing the currently sample-based counting of TP, FP and FN to bring the metrics closer to the intended use case of live predictions. For instance, TPs could be counted with regard to interlocks rather than samples, i.e. if at least one sample is positive in some range before an interlock, one TP is counted. It would also be reasonable to ignore consecutive FPs in a certain range or to ignore FP and FN occurring in a certain range before the interlock. 
Our customized target $T_s$ (beam time saved is constructed upon the definition of TP and FP. With the above adaptation, $T_s$ could also be brought closer to reality and become more instructive for real practices.

To enhance the model performance, one possible method is to tailor the training set according to model output. By excluding the interlocks that receive a low detection from the model, it might be possible to improve the detection of the remaining interlocks and reduce the number of FPs, thus increasing the overall model performance. 

Currently, a GUI for real-time implementation of the model has already been developed and tested on archived data. One of the next steps is to apply transfer learning to the model, train and update it with new data continuously, and eventually develop a real prognostic tool with GUI to be controlled by the machine operators.


%% file: Appendix.tex
\section{Interlock statistics} \label{apx:interlocks}
\begin{figure}[H]
    \centering
    \includegraphics[width=\linewidth]{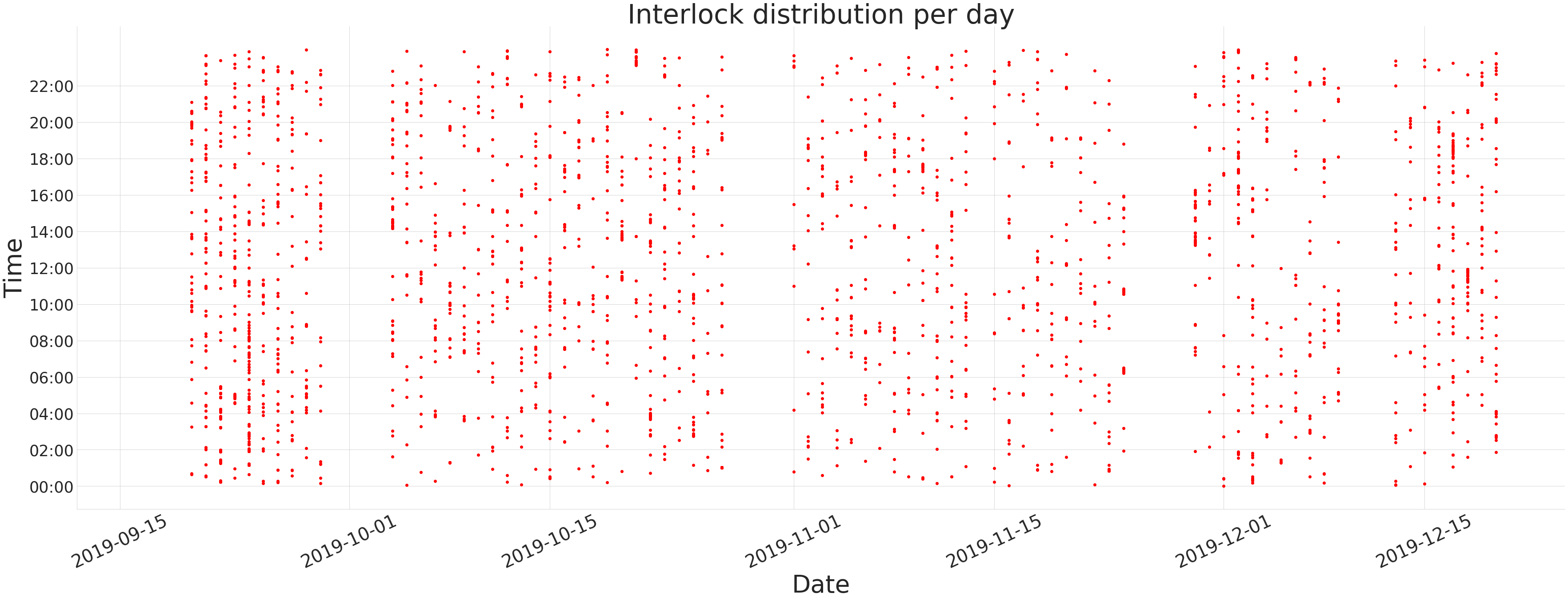}
    \caption{Time of all interlocks from 18.9.2019 to 25.12.2019. Regular maintenance periods can be observed.}
\label{fig:interlock_distribution}
\end{figure}

\begin{figure}[H]
    \centering
    \includegraphics[width=\linewidth]{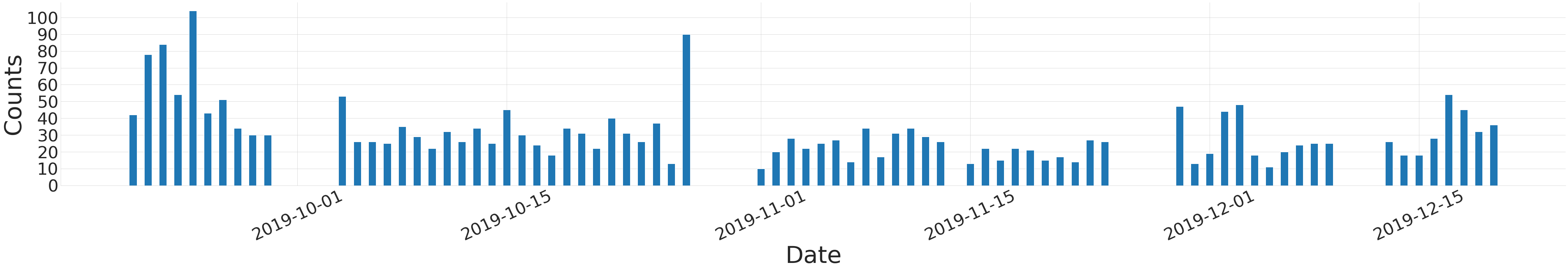}
    \caption{Number of interlocks happened per day from 18.9.2019 to 25.12.2019.}
\label{fig:interlock_count}
\end{figure}

\section{Example Recurrence Plots} \label{apx:recurrence}
\begin{figure}[H]
    \centering
    \includegraphics[width=0.75\linewidth]{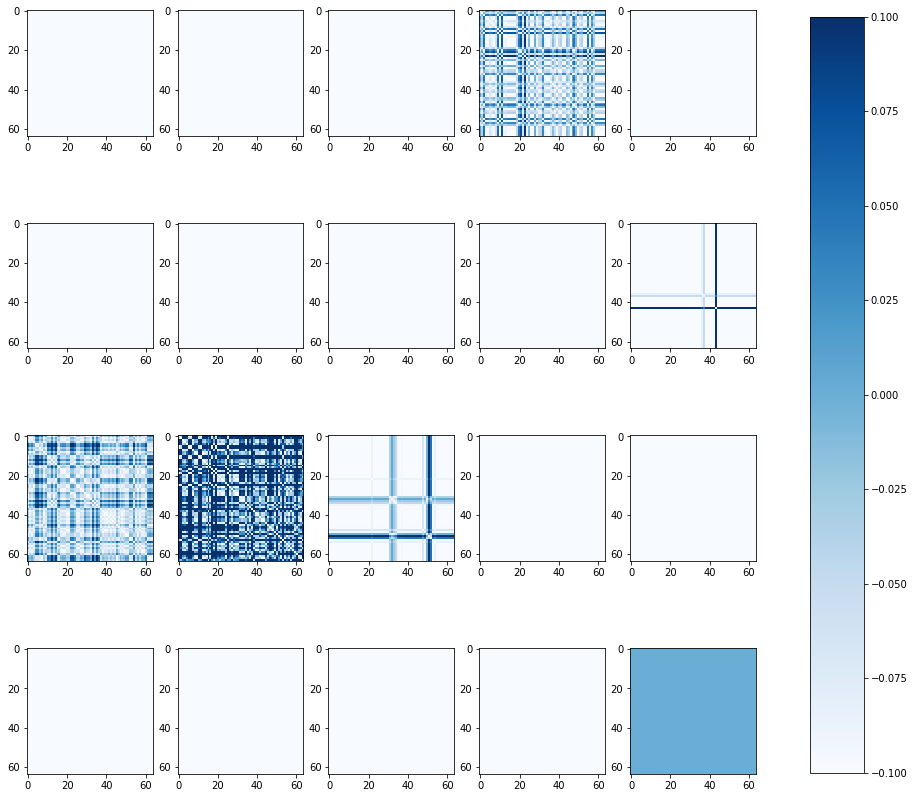}
\caption{The $20$ Recurrence Plots of an interlock sample.}
\label{fig:interlock_rp}
\end{figure}

\begin{figure}[H]
    \centering
    \includegraphics[width=0.75\linewidth]{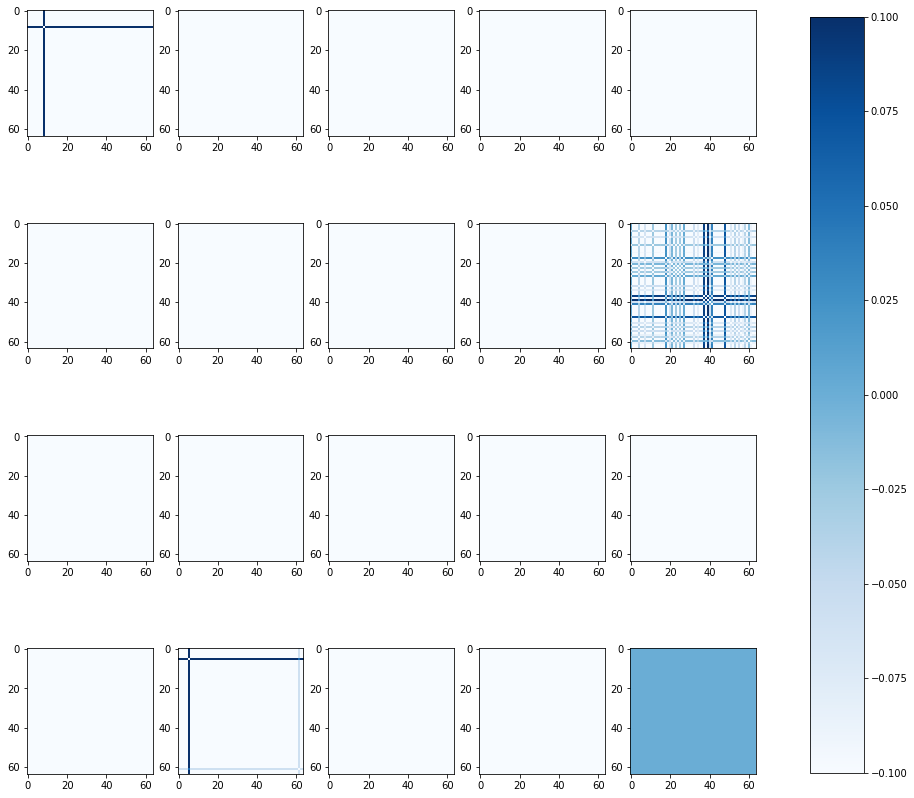}
\caption{The $20$ Recurrence Plots of a stable sample.}
\label{fig:stable_rp}
\end{figure}

\section{Architecture of the RPCNN model} \label{apx:rpcnn}

\begin{figure}[H]
    \centering
    \includegraphics[width=0.8\linewidth]{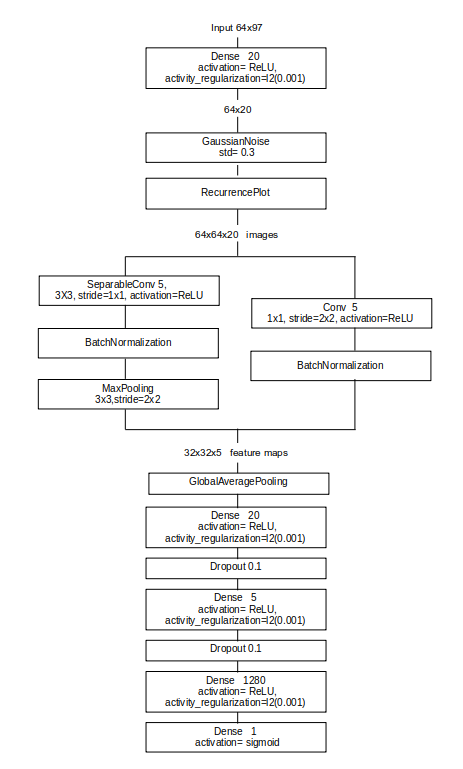}
\caption{The RPCNN architecture.}
\label{fig:model_architecture_details}
\end{figure}